\documentclass[epjST]{svjour}

\usepackage{graphics}

\usepackage{amsmath}

\usepackage{amssymb}

\newcommand\beq{\begin{equation}}
\newcommand\eeq{\end{equation}}
\newcommand\beqa{\begin{eqnarray}}
\newcommand\eeqa{\end{eqnarray}}
\newcommand{\nn}{\nonumber\\}

\newcommand{\zt}{\widetilde{\zeta}}
\newcommand{\at}{\widetilde{a}}
\newcommand{\et}{\epsilon}
\newcommand{\nuhs}{\nu_{\text{HS}}}

\newcommand{\al}{\alpha}

\newcommand{\dd}{\text{d}}

\newcommand{\NS}{\text{NS}}

\newcommand{\nutz}{\zeta}
\newcommand{\nuzt}{\nu_{0|2}}
\newcommand{\nuto}{\nu_{2|1}}
\newcommand{\nuzth}{\nu_{0|3}}

\newcommand{\omegazt}{\omega}

\newcommand{\omeganu}{\nu}

\begin{document}

\title{An exact solution of the inelastic Boltzmann equation for the Couette flow with uniform heat flux}

\author{Andr\'es Santos\thanks{\email{andres@unex.es}} \and Vicente Garz\'o\thanks{\email{vicenteg@unex.es}}
\and Francisco Vega Reyes\thanks{\email{fvega@unex.es}} }

\institute{Departamento de F\'{\i}sica, Universidad de
Extremadura, E-06071 Badajoz, Spain}

\abstract{
In the steady Couette flow of a granular gas the sign of the heat flux gradient is governed by the competition
between viscous heating and inelastic cooling. We show from the Boltzmann equation for inelastic Maxwell particles
that a special class of states exists where the viscous heating and the inelastic cooling exactly compensate each other at every point,
resulting in a uniform heat flux. In this state the (reduced) shear rate is enslaved to the coefficient of restitution $\alpha$,
so that the only free parameter is the (reduced) thermal gradient $\et$.
It turns out that the reduced moments of order $k$ are polynomials of degree $k-2$ in $\et$,
with coefficients that are nonlinear functions of $\alpha$.
In particular, the rheological properties ($k=2$) are independent of $\et$ and coincide exactly with those of the simple shear flow.
The heat flux ($k=3$) is linear in the thermal gradient (generalized Fourier's law),
but with an effective thermal conductivity differing from the Navier--Stokes one. In addition, a heat flux component parallel to the flow velocity and normal to the thermal gradient exists.
The theoretical predictions are validated by comparison with direct Monte Carlo simulations for the same model.
}

\maketitle

\section{Introduction}
\label{intro}

The influence of collisional dissipation on the dynamical properties of granular matter can be modeled by a fluid of inelastic hard spheres (IHS) with a constant coefficient of normal restitution $\alpha \leq 1$. For sufficiently low densities, the Boltzmann equation (BE) has been conveniently generalized \cite{BP04} to account for the inelasticity of binary collisions and the Navier--Stokes (NS) transport coefficients have been obtained in terms of the coefficient of restitution \cite{BDKS98,GM02,GD02} by means of the Chapman--Enskog method. However,  while the NS equations (constitutive equations that are linear in the hydrodynamic gradients) have been shown to be quite useful for describing several problems, in most situations of practical interest in granular gases (such as steady states) large gradients occur and more complex constitutive equations are required. This does not signal a breakdown of hydrodynamics, only a failure of the NS approximation.

One of the well-known examples of steady states is the simple or uniform shear flow (USF) problem \cite{Go03,SGD04}, characterized by a linear velocity field (i.e., $\partial u_x/\partial y=\text{const}$) and constant density $n$ and temperature $T$. This type of steady shear flow with zero heat flux can only occur when collisional cooling (which is fixed by the mechanical properties of the particles making up the granular gas) is exactly balanced by viscous heating (which is controlled by the shearing). Consequently, due to this intrinsic relationship between the shear field and dissipation, the (dimensionless) strength of the velocity gradient is set by the collisional cooling in the steady state.
This means that the corresponding hydrodynamic steady shear flow state is inherently non-Newtonian \cite{SGD04} (and so, beyond the scope of the NS equations) in  inelastic granular gases.

\begin{figure}
\resizebox{0.75\columnwidth}{!}{
\includegraphics{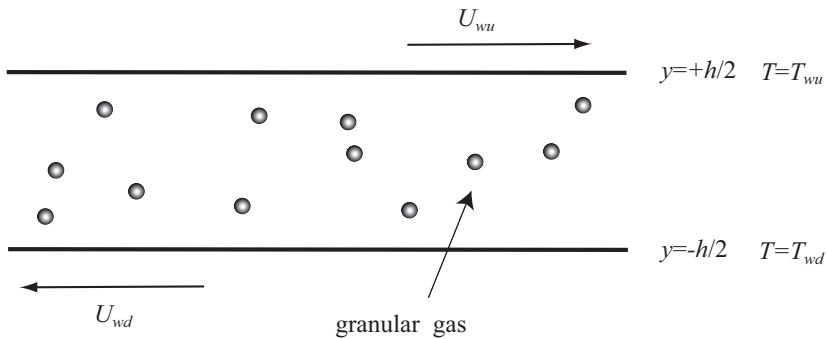}}
\caption{The planar Couette flow is driven by two horizontal plates, separated by a distance
$h$. Both act like sources of temperature and shear on a low density granular gas filling the space between them.} \label{setup}
\end{figure}

The steady planar Couette flow (cf.\ Fig.\ \ref{setup}) is much more complex than the USF. In contrast to the latter, the temperature is not uniform and thus a heat flux vector $\mathbf{q}$ coexists with the pressure tensor $P_{ij}$ \cite{TTMGSD01}. In general, inelastic cooling and viscous heating are unbalanced, their difference dictating the sign of the divergence of the heat flux \cite{TTMGSD01,VU09}. More explicitly, the energy balance equation reads
\begin{equation}
-\frac{\partial q_y}{\partial y}=\frac{d}{2}\zeta nT+P_{xy}\frac{\partial u_x}{\partial y},
\label{Tbal}
\end{equation}
where  $\zeta$ is the inelastic cooling rate and $d$ is the dimensionality of the system. Moreover, conservation of momentum implies $P_{xy}=\text{const}$ and $P_{yy}=\text{const}$.
Note that, while the inelastic cooling term $\zeta n T$ is inherently positive, the viscous heating term $P_{xy}{\partial u_x}/{\partial y}$ is inherently negative since the sign of the shear stress $P_{xy}$ is opposite to that of the shear rate $\partial u_x/\partial y$.

It can be shown that, for appropriate values of the boundary conditions for the velocity, both terms on the right-hand side of Eq.\ (\ref{Tbal}) can exactly balance at all points of the system \cite{VU09}.
This yields a whole new set of steady states  with a \emph{uniform} heat flux and other important common properties  \cite{VSG09}. For these flows, the streamwise heat flux component $q_x$ (absent at NS order) is also uniform and in general different from zero.
Of course, in the absence of thermal gradient, $q_y=0$ and one recovers the steady state condition defining the USF problem. On the other hand, in the elastic limit ($\alpha=1$ and $\zeta=0$)
but nonzero thermal gradient, the conventional Fourier flow description (where $\partial u_x/\partial y =0$, $P_{xy}=0$, and $q_x=0$)
for an ordinary gas is also recovered. Thus, the above two well-known steady states (inelastic USF and elastic Fourier flow) can be seen as particular situations of this class of steady Couette flows.
Apart from the condition $\mathbf{q}=\text{const}$, a unifying feature of this new class is that, while neither $u_x(y)$ nor $T(y)$ are linear, a parametric plot of $T$ vs $u_x$ shows a \emph{linear} relationship \cite{VU09,VSG09}. For this reason, we referred to this class of flows as LTu. The slope of the linear plot $T(u_x)$ goes from zero in the inelastic USF limit (constant temperature) to infinity in the elastic Fourier flow (zero macroscopic velocity).

Unfortunately, the analysis of the steady Couette flow from the BE for IHS is quite intricate and so evidence on the existence of the LTu class was provided in Ref.\ \cite{VSG09} by molecular dynamics simulations, a numerical solution of the BE by means of the direct  simulation Monte Carlo (DSMC) method \cite{B94},   and  an approximate solution obtained  by Grad's method. Of course, it is desirable to support the LTu class by means of analytical results cleanly derived from the BE. Like in the elastic case, a way to overcome such difficulties is by considering the so-called Maxwell models, i.e., models for which the collision rate is independent of the relative velocity of the two colliding particles \cite{BCG00,CCG00,NK00,C01,EB02a,EB02b,EB02c,NK03,G03,S03,SG07,G07,GS07,BC07,CCC09}. Thanks to this simplification, nonlinear transport properties can be \emph{exactly} obtained in some particular problems for elastic \cite{GS03,S09} and inelastic \cite{G03,SG07,G07} gases without the need of introducing additional, and sometimes uncontrolled, approximations. Apart from their academic interest, it has been also shown that in some cases the results derived from inelastic Maxwell models (IMM) compare well with those obtained from IHS \cite{G03,G07}.

In this paper we revisit the steady Couette flow with uniform heat flux (LTu class) in the framework of the BE for IMM. Given that the  shear rate and the coefficient of restitution are related through the condition $\partial q_y/\partial y=0$ in (\ref{Tbal}), the only nonequilibrium free parameter in the problem turns out to be the (reduced) thermal gradient $\et\propto \partial T/\partial y$. Based on  previous simulation and approximate results  obtained for IHS \cite{VSG09}, it is shown that the steady BE for IMM admits an {\em exact} solution characterized by constant pressure, $\partial u_x/\partial s = \text{const}$, and $\partial T/\partial s = \text{const}$, where $s$ is a conveniently scaled space variable. An analysis of the infinite hierarchy of velocity moments shows that the reduced moments  of order $k=2r+\ell$ are polynomials of parity $\ell$ and degree $k-2$ in $\et$, whose coefficients  are nonlinear functions of $\alpha$. The evaluation of the coefficients  associated with  the pressure tensor ($2r+\ell=2$) and with the heat flux ($2r+\ell=3$) is one of the main goals of this paper. The knowledge of the second-, third-, and fourth-order collisional moments for $d$-dimensional IMM \cite{GS07} allows us to explicitly get those coefficients. In particular, the expression for the pressure tensor coincides with the one previously obtained for the USF \cite{SG07}, while the heat flux is linear in the thermal gradient with two nonzero components ($q_y$ and $q_x$). The constitutive relation between $q_y$ and $\partial_y T$ can be seen as a generalized Fourier's law and defines an effective thermal conductivity different from the NS coefficient.
The existence of a component of the heat flux normal to the thermal gradient is a direct consequence of the inherent non-Newtonian features of the  Couette flow since this
component is absent in the NS description \cite{BDKS98}.

The above exact solution of the moment hierarchy resulting
 from the BE for IMM is free from boundary layers and thus it  holds in the bulk region of the granular fluid.
In  order to check its reliability under realistic boundary conditions, the BE for IMM has also been numerically solved  for a three-dimensional system by means of the DSMC method. The comparison between theory and simulation allows us to confirm that the analytical results are not mathematical artifacts but actually  describe the steady state achieved by the system in the bulk domain. As will be seen, the good agreement found between both approaches  clearly supports the validity of the assumptions made when working out the theoretical solution.

The plan of the paper is as follows. In Section \ref{sec2}, the BE for IMM is introduced and the explicit expressions for the collisional moments through third-order are given. Section \ref{sec3} deals with the problem we are interested in, namely the steady Couette flow with uniform heat flux  (LTu class). It is consistently proven that the hierarchy of moment equations associated with the problem admits a solution where the moments are polynomials in the thermal gradient with coefficients that are nonlinear functions of the coefficient of restitution. Next, the explicit expressions for the momentum and heat fluxes are obtained in Section \ref{sec4}, while the comparison with Monte Carlo simulations is presented in Section \ref{sec5}. The paper is closed in Section \ref{sec6} with a brief discussion of the results derived here.

\section{The inelastic Maxwell model}
\label{sec2}

Let us consider a granular fluid modeled as a gas of inelastic Maxwell particles.
A constant parameter, the coefficient of normal restitution  $\alpha$, accounts for the inelasticity in the collisions. Its values range from $\alpha=0$ (purely inelastic collision) to $\alpha=1$ (purely elastic collision). At a kinetic-theory level, all the relevant information on the state of the system is conveyed by the one-particle velocity distribution function $f(\mathbf{r},\mathbf{v};t)$. In the low-density regime, the inelastic BE for IMM reads
\beq
\left(\partial_t
+\mathbf{v}\cdot\nabla\right)f(\mathbf{r},\mathbf{v};t)=J[\mathbf{v}|f,f],
\label{2.1}
\eeq
where the Boltzmann collision operator $J[\mathbf{v}|f,f]$ is given
by \cite{BCG00,NK00,EB02a,EB02b,EB02c,NK03}
\begin{equation}
J\left[{\bf v}_{1}|f,f\right] =\frac{\omeganu}{n\Omega_d} \int
\dd{\bf v}_{2}\int \dd\widehat{\boldsymbol{\sigma}} \left[
\alpha^{-1}f({\bf v}_{1}')f({\bf v}_{2}')-f({\bf v}_{1})f({\bf
v}_{2})\right] .
\label{1}
\end{equation}
Here,
\beq
n=\int\dd \mathbf{v} f(\mathbf{v})
\eeq
is the number density, $\omeganu$ is the collision frequency
(assumed to be independent of $\alpha$),
$\Omega_d=2\pi^{d/2}/\Gamma(d/2)$ is the total solid angle in $d$
dimensions, and the primes on the velocities denote the
initial values $\{{\bf v}_{1}^{\prime}, {\bf v}_{2}^{\prime}\}$ that
lead to $\{{\bf v}_{1},{\bf v}_{2}\}$ following a binary collision:
\begin{equation}
\label{3b}
{\bf v}_{1}^{\prime}={\bf v}_{1}-\frac{1}{2}\left( 1+\alpha
^{-1}\right)(\widehat{\boldsymbol{\sigma}}\cdot {\bf
g})\widehat{\boldsymbol {\sigma}}, \quad {\bf v}_{2}^{\prime}={\bf
v}_{2}+\frac{1}{2}\left( 1+\alpha^{-1}\right)
(\widehat{\boldsymbol{\sigma}}\cdot {\bf
g})\widehat{\boldsymbol{\sigma}}\;,
\end{equation}
where ${\bf g}={\bf v}_1-{\bf v}_2$ is the relative velocity of the
colliding pair and $\widehat{\boldsymbol{\sigma}}$ is a unit vector
directed along the centers of the two colliding particles.

Apart from $n$, the basic moments of $f$ are the flow velocity
\beq
\mathbf{u}=\frac{1}{n}\int\dd \mathbf{v} \mathbf{v}f(\mathbf{v})
\eeq
and the granular temperature
\beq
T=\frac{m}{dn}\int\dd \mathbf{v}\, V^2 f(\mathbf{v}),
\label{granT}
\eeq
where $\mathbf{V}=\mathbf{v}-\mathbf{u}(\mathbf{r})$ is the peculiar
velocity. The momentum and energy fluxes are characterized by the
pressure tensor
\beq
P_{ij}=m\int\dd \mathbf{v} \,V_i V_j f(\mathbf{v})
\label{Pij}
\eeq
and the heat flux
\beq
\mathbf{q}=\frac{m}{2}\int\dd \mathbf{v}\,
V^2\mathbf{V}f(\mathbf{v}),
\label{qi}
\eeq
respectively. Finally, the rate of energy dissipated due to collisions defines the
cooling rate $\zeta$ as
\beq
\zeta=-\frac{m}{dnT}\int\dd \mathbf{v}\, V^2 J[\mathbf{v}|f,f].
\label{zeta}
\eeq

The collision frequency $\nu$ is proportional to density and, in general, is also a function of the granular temperature $T$. In fact, in order to make contact with the IHS model, one must take $\nu\propto n T^{1/2}$ \cite{BCG00}. In particular, the choice \cite{S03}
\begin{equation}
\label{4b}
\omeganu=\frac{d+2}{2}\nuhs, \quad
\nuhs=\frac{4\Omega_d}{\sqrt{\pi}(d+2)}n\sigma^{d-1}\sqrt{\frac{T}{m}},
\end{equation}
where $\sigma$ is the diameter of the spheres, yields the same expression for the cooling rate as the one found for IHS
(evaluated in the local equilibrium approximation). This will be the choice made here, although most of the results are actually independent of the explicit form of $\nu$.
The collision frequency $\nuhs$ is the one associated with the NS shear viscosity of an ordinary gas  ($\alpha=1$) of hard spheres, i.e., $\eta_{\NS}=p/\nuhs$.

The main advantage of the BE for Maxwell
models (both elastic and inelastic) is that the (collisional)
moments of the operator $J[f,f]$ can be exactly evaluated in terms of the moments of
$f$, without the explicit knowledge of the latter \cite{TM80}.  In particular, the second- and third-order collisional moments are \cite{S03,GS07}
\beq
m\int\dd \mathbf{V} \, V_iV_j J[\mathbf{V}|f,f]=-\nuzt(P_{ij}-p\delta_{ij})-\nutz p \delta_{ij},
\label{Z1}
\eeq
\beq
\frac{m}{2}\int\dd \mathbf{V} \, V_iV_jV_k J[\mathbf{V}|f,f]=-\nuzth Q_{ijk}-\frac{\nuto-\nuzth}{d+2}\left(q_i\delta_{jk}+q_j\delta_{ik}+q_k\delta_{ij}\right),
\label{Z2}
\eeq
where $p=nT=\frac{1}{d}\text{tr}\,\mathsf{P}$ is the hydrostatic pressure,
\beq
Q_{ijk}=\frac{m}{2}\int \dd\mathbf{V}\, V_iV_jV_k f(\mathbf{V})
\label{Z3}
\eeq
is a third-rank tensor, and
\beq
\nutz=\frac{1-\al^2}{2d}\nu,
\label{X6ab}
\eeq
\beq
\nuzt=\nutz+\frac{(1+\alpha)^2}{2(d+2)}\nu,\quad \nuto=\frac{3}{2}\nutz+\frac{(1+\al)^2(d-1)}{2d(d+2)}\nu,\quad \nuzth=\frac{3}{2}\nuzt.
\label{X6}
\eeq
In Eq.\ \eqref{X6} the collision frequencies $\nuzt$ and $\nuto$  are decomposed into a part
inherent to the collisional cooling plus the genuine part associated with  the
momentum and energy collisional transfers. For later use, it is convenient to define
\beqa
\omegazt&\equiv& \nuzt-\zeta=\frac{(1+\alpha)^2}{2(d+2)}\nu\nn
&=&\left(\frac{1+\al}{2}\right)^2\nuhs.
\label{Z4}
\eeqa
Thus the collision frequency $\nuhs$ defined by the first equality of Eq.\ \eqref{4b} is actually the value of $\omegazt$ in the elastic limit $\al\to 1$.

The fourth-order collisional moments were also explicitly evaluated in Ref.\ \cite{GS07} but will not be displayed here \cite{note}.

\section{Couette flow with uniform heat flux}
\label{sec3}

The planar Couette flow considered in this paper corresponds to a
granular gas enclosed between two parallel, infinite plates (normal to the $y$ axis) at $y=\pm h/2$ in relative motion along the $x$ direction, and kept at different temperatures (cf.\ Fig.\ \ref{setup}). The resulting flow velocity is along the $x$ axis and, from symmetry, it is expected that the hydrodynamic fields only vary in the $y$ direction. Consequently,  the velocity distribution function is also assumed to depend  on the coordinate $y$ only. Moreover, we focus on the steady state, so Eq.\ \eqref{2.1} becomes
\beqa
v_y\partial_s f&=&\frac{1}{\omegazt}J[f,f]\nn
&=&\frac{1}{n\Omega_d}\frac{2(d+2)}{(1+\alpha)^2} \int
\dd{\bf v}_{2}\int \dd\widehat{\boldsymbol{\sigma}} \left[
\alpha^{-1}f({\bf v}_{1}')f({\bf v}_{2}')-f({\bf v}_{1})f({\bf
v}_{2})\right],
\label{3}
\eeqa
where we have introduced the scaled variable $s$ as
\beq
\dd s=\omegazt \dd y.
\label{2}
\eeq
Note that Eq.\ \eqref{3} is ``universal'' in the sense that it is independent of the precise choice of $\nu$.

Now we assume that an exact solution of Eq.\ \eqref{3} exists characterized by
\beq
p=\text{const},\quad \frac{\partial u_x}{\partial s}=\at=\text{const},\quad \frac{\partial T}{\partial s}=\text{const}.
\label{4}
\eeq
The constant $\at$  represents a Knudsen number associated with the shear rate. There is another Knudsen number associated with the thermal gradient, namely
\beq
\et=\sqrt{2T/m}\frac{\partial \ln T}{\partial s}.
\label{5}
\eeq
This quantity is not constant since $\partial_s T=\text{const}$ implies $\et\propto T^{-1/2}$. {}From Eqs.\ \eqref{4} and \eqref{5} we get
\beq
\frac{\partial T}{\partial u_x}=\frac{\et\sqrt{mT/2}}{\at}=\text{const}.
\label{LTu}
\eeq
This means that when the spatial variable ($y$ or $s$) is eliminated to express $T$ as a function of $u_x$ one gets a linear relationship. Therefore, the solution to the BE \eqref{3} consistent with the hydrodynamic profiles \eqref{4} defines the LTu class of Couette flows described in the Introduction.

 As we will see, the consistency of the profiles \eqref{4} is possible only if $\at$ takes a special value for each coefficient of restitution $\al$. In contrast, the reduced thermal gradient $\et$ will remain free and so independent of $\al$.

In terms of the  variable $s$, the temperature profile is
\beq
T(s)=T_0\left(1+\frac{\et_0}{v_0}s\right),\quad v_0\equiv\sqrt{2T_0/m},
\label{19}
\eeq
where $T_0$ and $\et_0$ are the temperature and Knudsen number at a reference point $s=0$.
In order to get the dependence of temperature in real space, we need to make use of the precise temperature dependence of $\omegazt$. According to the assumption $\omegazt\propto n T^{1/2}=p T^{-1/2}$ [see Eq.\ \eqref{4b}], Eq.\ \eqref{2} implies
\beqa
y(s)&=&y_0+\frac{1}{\omegazt_0}\int_0^s \dd s'\, \sqrt{T(s')/T_0}\nn
&=&y_0+\frac{2v_0}{3\et_0\omegazt_0}\left[\left(T/T_0\right)^{3/2}-1\right],
\label{20}
\eeqa
where  we have  called $\omegazt_0$ to the value of $\omegazt$  at $s=0$.
Inversion of Eq.\ \eqref{20} yields
\beq
T(y)=T_0\left[1+\frac{3\et_0\omegazt_0}{2v_0}(y-y_0)\right]^{2/3}.
\label{21}
\eeq
Therefore, $T^{3/2}$ is a linear function of $y$. Note that $\et\propto \partial T/\partial y$:
\beq
\et=\frac{v_0}{\omegazt_0T_0}\frac{\partial T}{\partial y}.
\label{24}
\eeq
Equations \eqref{20}--\eqref{24} are only valid if $\omegazt\propto n T^{1/2}$ (Maxwell particles mimicking hard spheres), while Eqs.\ \eqref{4}--\eqref{19} are ``universal'' in the sense described above. In the remainder of this section we will not need to specify the temperature dependence of the collision frequencies.

We now introduce the dimensionless velocity distribution function
\beq
\varphi(\mathbf{c};\et)=\frac{T(s)}{p}\left[\frac{2T(s)}{m}\right]^{d/2}f(s,\mathbf{v}),\quad \mathbf{c}=\frac{\mathbf{v}-\mathbf{u}(s)}{\sqrt{2T(s)/m}}.
\label{6}
\eeq
Upon writing $\varphi(\mathbf{c};\et)$ we have not included the dependence on the reduced shear rate $\at$ since, as said before, it is not a free parameter in this solution but it is enslaved to the value of $\alpha$.
Since, according to Eq.\ \eqref{6}, all the dependence of $f$ on $s$ occurs through the hydrodynamic fields $T$ and $u_x$, we have
\beq
\frac{\partial f}{\partial s}=\frac{\partial T}{\partial s}\frac{\partial f}{\partial T}+ \frac{\partial u_x}{\partial s}\frac{\partial f}{\partial u_x},
\label{7}
\eeq
where
\beq
\frac{\partial f}{\partial u_x}=-\frac{p}{T}\left(\frac{m}{2T}\right)^{(d+1)/2}\frac{\partial \varphi}{\partial c_x},
\label{8}
\eeq
\beq
\frac{\partial f}{\partial T}
=\frac{p}{T} \left(\frac{m}{2T}\right)^{d/2}\left(-\frac{d+2}{2} T^{-1}\varphi+\frac{\partial \mathbf{c}}{\partial T}\cdot \frac{\partial \varphi}{\partial \mathbf{c}}
+\frac{\partial \et}{\partial T} \frac{\partial \varphi}{\partial \et}\right).
\label{9}
\eeq
Taking into account that $\partial \mathbf{c}/\partial T=-\frac{1}{2}T^{-1}\mathbf{c}$ and  $\partial \et/\partial T=-\frac{1}{2}T^{-1}\et$, one finally gets
\beq
\frac{\partial}{\partial s}
f(s,\mathbf{v})=-\frac{p}{T}\left(\frac{m}{2T}\right)^{(d+1)/2}\left[\frac{\et}{2}\left(2+
\frac{\partial}{\partial\mathbf{c}}\cdot\mathbf{c}+\et\frac{\partial}{\partial
\et}\right)+\at\frac{\partial}{\partial c_x}\right]\varphi(\mathbf{c};\at,\et).
\label{10}
\eeq
Consequently,  Eq.\ \eqref{3} becomes
\beqa
-c_y\left[\frac{\et}{2}\left(2+
\frac{\partial}{\partial\mathbf{c}}\cdot\mathbf{c}+\et\frac{\partial}{\partial
\et}\right)+\at\frac{\partial}{\partial c_x}\right]\varphi&=&\frac{2(d+2)}{(1+\alpha)^2\Omega_d}\int
\dd{\bf c}_{1}\int \dd\widehat{\boldsymbol{\sigma}} \left[
\alpha^{-1}\varphi({\bf c}')\varphi({\bf c}_{1}')-\varphi({\bf c})\varphi({\bf c}_{1})\right] \nn
&\equiv &\mathcal{J}[\mathbf{c}|\varphi,\varphi].
\label{11}
\eeqa

We consider now the following  moments of order $k=2r+\ell$,
\beq
\mathcal{M}_{2r|\ell,{h}}(\et)=\int \dd\mathbf{c}\,c^{2r}c_y^{\ell{-h}}{c_x^h}
\varphi(\mathbf{c};\et)\equiv \langle \,c^{2r}c_y^{\ell{-h}}{c_x^h}\rangle,\quad 0\leq h\leq\ell.
\label{12}
\eeq
By definition, $\mathcal{M}_{0|0,0}=1$, $\mathcal{M}_{0|1,0}=\mathcal{M}_{0|1,1}=0$, and $\mathcal{M}_{2|0,0}=\frac{d}{2}$. According to Eq.\ \eqref{11}, the moment equations read
\beqa
\frac{\et}{2}\left(2r+\ell-1-\et\frac{\partial}{\partial
\et}\right)\mathcal{M}_{2r|\ell+1,{h}}{+\at\left(2r \mathcal{M}_{2r-2|\ell+2,h+1}+h\mathcal{M}_{2r|\ell,h-1}\right)}
&=&\mathcal{J}_{2r|\ell,{h}},
\label{13}
\eeqa
where
\beq
\mathcal{J}_{2r|\ell,{h}}\equiv\int \dd\mathbf{c}\,c^{2r}c_y^{\ell{-h}}{c_x^h} \mathcal{J}[\mathbf{c}|\varphi,\varphi]
\eeq
are the corresponding collisional moments. As we are considering Maxwell models, $\mathcal{J}_{2r|\ell,{h}}$ are given as bilinear combinations of the form  $\mathcal{M}_{2r'|\ell',{h'}}\mathcal{M}_{2r''|\ell'',{h''}}$ such that $2r'+\ell'+2r''+\ell''=2r+\ell$. Therefore, only moments of order  equal to or smaller than $2r+\ell$ contribute to $\mathcal{J}_{2r|\ell,{h}}$.
In particular, from Eqs.\ \eqref{Z1} and \eqref{Z2} we have \cite{GS07}
\beq
\mathcal{J}_{2|0,0}=-\zt\frac{d}{2},
\label{Z5}
\eeq
\beq
\mathcal{J}_{0|2,h}=-\left(1+\zt\right)\mathcal{M}_{0|2,h}+\begin{cases}
  \frac{1}{2},& h=0,2\\
  0,& h=1,
\end{cases},
\label{Z6}
\eeq
\beq
\mathcal{J}_{2|1,h}=-\left(\frac{d-1}{d}+\frac{3}{2}\zt\right)\mathcal{M}_{2|1,h},\quad h=0,1,
\label{Z7}
\eeq
\beq
\mathcal{J}_{0|3,h}=-\frac{3}{2}\left(1+\zt\right)\mathcal{M}_{0|3,h}+\frac{1}{2d}\times\begin{cases}
  3 \mathcal{M}_{2|1,0},&h=0,\\
   \mathcal{M}_{2|1,1},&h=1,\\
    \mathcal{M}_{2|1,0},&h=2,\\
    3 \mathcal{M}_{2|1,1},&h=3.
\end{cases}
\label{Z8}
\eeq
Here,
\beq
\zt\equiv\frac{\nutz}{\omegazt}=\frac{d+2}{d}\frac{1-\al}{1+\al},
\label{X6a}
\eeq
and we have taken into account that
\beq
\frac{\nuzt}{\omegazt}=1+\zt, \quad \frac{\nuto}{\omegazt}=\frac{d-1}{d}+\frac{3}{2}\zt, \quad
 \frac{\nuzth}{\omegazt}=\frac{3}{2}(1+\zt).
 \label{X7}
 \eeq

\begin{figure}
\resizebox{0.75\columnwidth}{!}{%
  \includegraphics{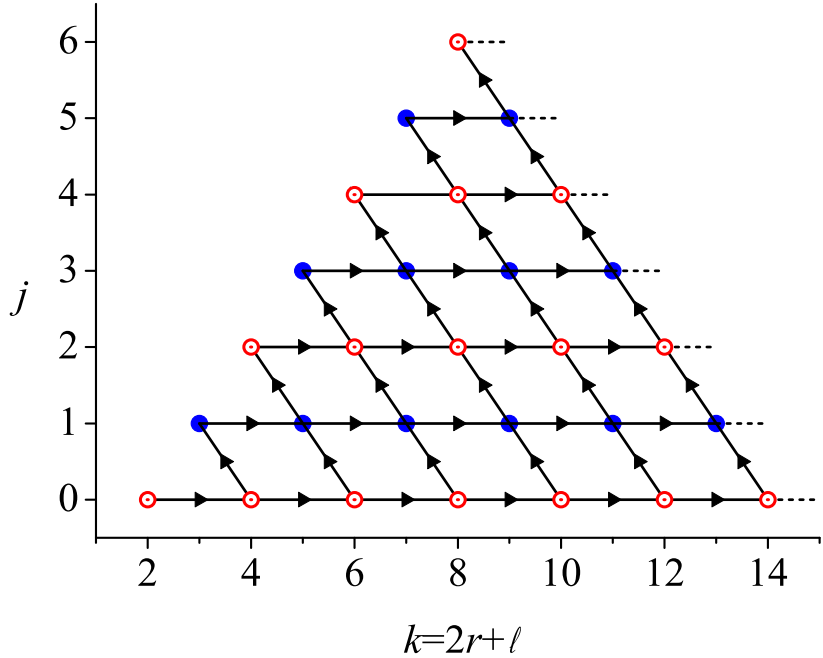} }
\caption{Sketch of the sequence followed in the recursive determination of the  coefficients $\mu_j^{(2r|\ell,h)}$. All the coefficients of the same order $k=2 r+\ell$ are represented by a common circle. The zeroth-degree coefficients $\mu_0^{(2r|\ell,h)}$ correspond to the USF.}
\label{arrows}
\end{figure}

Let us check that the hierarchy \eqref{13} is consistent with solutions of the form
\beq
\mathcal{M}_{2r|\ell, h}(\et)=\sum_{j=0}^{2r+\ell-2}\mu_j^{(2r|\ell, h)}\et^j,\quad \mu_j^{(2r|\ell. h)}=0\text{ if }j+\ell=\text{odd},
\label{14}
\eeq
i.e., the moments $\mathcal{M}_{2r|\ell,{h}}(\et)$ of order $2r+\ell\geq 2$ are  \emph{polynomials} in the thermal Knudsen number $\et$ of degree $2r+\ell-2$ and parity $\ell$.
 First note that  the moment $\mathcal{M}_{2r|\ell+1,h}$ in the first term on the left-hand side of Eq.\ \eqref{13} is a polynomial of degree $2r+\ell-1$ and has a parity different from that of $\mathcal{M}_{2r|\ell, h}$. However, the action of the operator $\et(2r+\ell-1-\et\partial/\partial\et)$ restores the parity and transforms the degree of the polynomial into $2r+\ell-2$. The second term on the left-hand side obviously has the same parity and degree as $\mathcal{M}_{2r|\ell, h}$. Finally, the condition $2r'+\ell'+2r''+\ell''=2r+\ell$ assures that the products $\mathcal{M}_{2r'|\ell',h'}\mathcal{M}_{2r''|\ell'',h''}$ are polynomials of the same degree and parity as those of $\mathcal{M}_{2r|\ell, h}$.

 It is worthwhile noticing that the hierarchy \eqref{13} reduces to that of the USF problem for IMM when $\et=0$ \cite{SG07}, while it reduces to that of the conventional Fourier flow problem for elastic Maxwell particles when $\al=1$ (which, as will be seen below, implies $\at=0$) \cite{S09}.
  The general problem ($\et\neq 0$, $\at\neq 0$) is much more difficult since it combines both momentum and energy transport. However, the moment hierarchy  can be exactly solved via a recursive scheme. The key point is that, while Eq.\ \eqref{13} includes the moment $\mathcal{M}_{2r|\ell+1,h}$, which is a polynomial  of degree $2r+\ell-1$, only the coefficients $\mu_j^{(2r|\ell+1,h)}$ with $j\leq 2r+\ell-3$ contribute to Eq.\ \eqref{13}. Therefore, one can compute the independent coefficients $\mu_0^{(2r|\ell,h)}$ (which are in fact the only ones appearing in the USF), then the first-degree coefficients $\mu_1^{(2r|\ell,h)}$, and so on.
 The recursive scheme is sketched in Fig.\ \ref{arrows}, where the arrows indicate the sequence followed in the determination of $\mu_j^{(2r|\ell,h)}$. The open (closed) circles represent the coefficients associated with even (odd) $j$ and $\ell$. Notice that the number of coefficients needed to determine a given moment $\mathcal{M}_{2r|\ell,h}$ is finite. In fact, the coefficients represented in Fig.\ \ref{arrows} are the ones involved in the evaluation of $\mathcal{M}_{2r|\ell,h}$ for $k=2r+\ell\leq 8$.
 In the next Section we follow this scheme to determine the pressure tensor ($k=2$) and the heat flux ($k=3$).

It must be noted that, although all the moments can in principle be evaluated recursively, the explicit form of the reduced distribution function $\varphi(\mathbf{c};\et)$ is not known. However, one could take advantage of the knowledge of the first few moments to construct an \emph{approximate} distribution function compatible with them by using, for instance, maximum-entropy arguments.

Before closing this section, let us  consider the spatial dependence of the \emph{dimensional} moments
\beqa
{M}_{2r|\ell,{h}}(y)&=&\int \dd\mathbf{v}\,V^{2r}V_y^{\ell{-h}}{V_x^h}
f(y,\mathbf{v})\nn
&=& \frac{p}{T(y)} \left[\frac{2T(y)}{m}\right]^{r+\ell/2}\mathcal{M}_{2r|\ell,h}(\et).
\label{3.15}
\eeqa
Taking into account Eq.\ \eqref{14} and the fact that $\sqrt{T(y)}\et(y)=\sqrt{T_0}\et_0=\text{const}$, one can easily get
\beq
M_{2r|\ell,{h}}(y)=\frac{2p}{m}\left(\et_0^2\frac{2T_0}{m}\right)^{r+\ell/2-1}\,\sum_{j=0}^{[(2r+\ell-2)/2]}
\mu_{2r+\ell-2-2j}^{(2r|\ell,h)}\left[\frac{T(y)}{\et_0^2T_0}\right]^j,
\label{3.16}
\eeq
where here $[x]$ denotes the integer part of $x$ and it is understood that $2r+\ell\geq 2$. Equation \eqref{3.16} shows that $M_{2r|\ell,{h}}(y)$ is just a polynomial in $T(y)$ of degree $[(2r+\ell-2)/2]$. In particular, the second- and third-order moments are uniform, the fourth- and fifth-order moments depend on $y$ via a linear dependence on $T(y)$, and so on.
The spatial uniformity of the third-order moments implies that the LTu solution characterized by the hydrodynamic profiles \eqref{4} corresponds indeed to planar Couette flows with uniform heat flux.

\section{Pressure tensor and heat flux}
\label{sec4}
Let us start considering the moments of second order (pressure tensor). This will give us the dependence of the reduced shear rate $\at$ and the rheological properties on dissipation.

Taking $(2r|\ell,h)=(2|0,0)$, $(0|2,0)$, $(0|2,1)$, and $(0|2,2)$ in Eq.\ \eqref{13}, one gets a coupled set of linear equations for $\at$ and   $\mathcal{M}_{0|2,h}$ ($h=0,1,2$). The solution is
\beq
\at=\sqrt{\frac{d\zt}{2}}(1+\zt),
\label{15}
\eeq
\beq
\mathcal{M}_{0|2,1}=\langle c_x c_y\rangle=-\frac{\at}{2(1+\zt)^2},
\label{16}
\eeq
\beq
\mathcal{M}_{0|2,0}=\langle c_y^2\rangle=\frac{1}{2(1+\zt)},
\label{17}
\eeq
\beq
\mathcal{M}_{0|2,2}=\langle c_x^2\rangle=\frac{1+d\zt}{2(1+\zt)}.
\label{18}
\eeq
These quantities depend on the coefficient of restitution $\al$ through the scaled cooling rate $\zt$ given by Eq.\ \eqref{X6a}.

Next, we consider the third-order moments. Setting $(2r|\ell,h)=(2|1,0)$, $(2|1,1)$, $(0|3,0)$, $(0|3,1)$, $(0|3,2)$, and $(0|3,3)$ in Eq.\ \eqref{13}, we obtain a set of six linear equations whose solution gives the third-order moments in terms of the independent terms of the fourth-order moments and $\zt$. Here we only display the explicit forms for the two third-order moments defining the $x$ and $y$ components of the heat flux. They are
\beqa
\mathcal{M}_{2|1,0}(\et)&=&\langle c^2 c_y\rangle=-\frac{2d\et}{ X}\bigg\{
  4 d  (2d - 2 + 3d \zt) \zt \mu_0^{(0|4,0)}+
      8 d\zt \mu_0^{(0|4,2)} + (18 d -18+ 19d \zt)\mu_0^{(2|2,0)}\nn
      &&\ -
  6 \sqrt{2d\zt}  \left[ (2d - 2 + 3d \zt) \mu_0^{(0|4,1)} +
     \mu_0^{(2|2,1)}\right]\bigg\} ,
  \label{Z9}
       \eeqa
 \beqa
\mathcal{M}_{2|1,1}(\et)&=&\langle c^2 c_x\rangle
=\frac{2d\et}{3X}   \bigg\{\sqrt{2d\zt}\big[4 d (7d- 2  + 9 d \zt) \zt\mu_0^{(0|4,0)} +
 6  (6 d - 6 + 5 d\zt) \mu_0^{(0|4,2)}\nn&& +
 3 (9{d} + 6 + 17 d\zt) \mu_0^{(2|2,0)}\big]-12 d \zt  (7d - 2  + 9 d \zt) \mu_0^{(0|4,1)}\nn
 && -
 9  (6 d - 6 + 5d \zt) \mu_0^{(2|2,1)}\bigg\},
 \label{Z10}
       \eeqa
 where
 \beq
 X\equiv  36 (d - 1)^2 -  d\left(76 -56 d - 9  d\zt\right)\zt.
       \label{Z11}
 \eeq
Thus, in order to complete the determination of $\mathcal{M}_{2|1,0}$ and $\mathcal{M}_{2|1,1}$ we need to know the $\alpha$-dependence of the coefficients $\mu_0^{(0|4,0)}$, $\mu_0^{(0|4,2)}$, $\mu_0^{(2|2,0)}$, $\mu_0^{(0|4,1)}$, and $\mu_0^{(2|2,1)}$. These coefficients can be obtained from Eq.\ \eqref{13} by taking $\et=0$. This was in fact already done in Ref.\ \cite{SG07} in the context of the solution of the USF for IMM. The final expressions for the three-dimensional case ($d=3$) are
\beq
\frac{\mathcal{M}_{2|1,0}(\et)}{\et}=-\frac{135 (1 + \al)^2}{ 8  (4 - \al)^2 (829 - 162 \al - 91 \al^2)}\frac{A(\al)}{C(\al)},
\label{Z17}
\eeq
\beq
\frac{\mathcal{M}_{2|1,1}(\et)}{\et}=\frac{27\sqrt{5/2} (1 + \al)\sqrt{1-\al^2}}{ 4  (4 - \al)^2 (829 - 162 \al - 91 \al^2)}\frac{B(\al)}{C(\al)},
\label{Z18}
\eeq
where
\beq
A(\al)=\sum_{i=0}^{26} A_i (1-\al)^i,\quad B(\al)=\sum_{i=0}^{26} B_i (1-\al)^i, \quad C(\al)=\sum_{i=0}^{24} C_i (1-\al)^i.
\label{Z19}
\eeq
The numerical values of the coefficients $A_i$, $B_i$, and $C_i$ are given in Table \ref{tab:1}.

\begin{table}
\caption{Numerical values of the coefficients $A_i$, $B_i$, and $C_i$ defined in Eq.\ \protect\eqref{Z19}.}
\label{tab:1}       
\begin{tabular}{llll}
\hline\noalign{\smallskip}
$i$ & $A_i$& $B_i$& $C_i$  \\
\noalign{\smallskip}\hline\noalign{\smallskip}
 $0 $&$ 341218675233128448 $&$ 2985663408289873920 $&$ 2369574133563392 $\\
 $1 $&$ 2482013713849384960 $&$ 20450944454204522496 $&$ 7160814742552576 $\\
 $2 $&$ 7969339170892868608 $&$ 58568421476117803264 $&$ 5980987758379200 $\\
 $3 $&$ 14758809962784485248 $&$ 96685932126035252672 $&$ -1622483991362528 $\\
 $4 $&$ 17947525965827798144 $&$ 107159766670778436160 $&$ -6127404964857736 $\\
 $5 $&$ 15792893280853221008 $&$ 88222721223517916056 $&$ -5323957544763312 $\\
 $6 $&$ 10848492042169821200 $&$ 57714899464650979880 $&$ -3091956803616730 $\\
 $7 $&$ 6083138827020141820 $&$ 31096845798718075642 $&$ -1463253462204164 $\\
 $8 $&$ 2852642631673202332 $&$ 14152867760388474114 $&$ -595099812423405 $\\
 $9 $&$ 1160168028460008358 $&$ 5649360969446546811 $&$ -211065154772581 $\\
 $10 $&$ 424875366505855228 $&$ 2030043005756227200 $&$ -63304902022304 $\\
 $11 $&$ 137908868657776387 $&$ 642139536452566558 $&$ -15813143468850 $\\
 $12 $&$ 37996059299305384 $&$ 172739779638915362 $&$ -3363854349023 $\\
 $13 $&$ 8835881256127730 $&$ 39251245670664391 $&$ -521942377705 $\\
 $14 $&$ 1676141413944428 $&$ 7120337713266190 $&$ -23351656362 $\\
 $15 $&$ 194128153036217 $&$ 724478043152468 $&$ 14744389802 $\\
 $16 $&$ -14258415439016 $&$ -96248350169696 $&$ 7223865310 $\\
 $17 $&$ -15161725120946 $&$ -74193692948568 $&$ 2126089824 $\\
 $18 $&$ -5457337268162 $&$ -25236538926270 $&$ 429672915 $\\
 $19 $&$ -1406011921458 $&$ -6243932259627 $&$ 78316371 $\\
 $20 $&$ -274206254448 $&$ -1200674292216 $&$ 11377737 $\\
 $21 $&$ -47059936911 $&$ -201709797345 $&$ 1142559 $\\
 $22 $&$ -6553863486 $&$ -27252518922 $&$ 141858 $\\
 $23 $&$ -661870251 $&$ -2792296044 $&$ 5508 $\\
 $24 $&$ -75802878 $&$ -300657636 $&$ 648 $\\
 $25 $&$ -3147012 $&$ -13199760 $&$ 0 $\\
 $26 $&$ -329832 $&$ -1263600 $&$ 0 $\\
\noalign{\smallskip}\hline
\end{tabular}
\end{table}

Once the pressure tensor and the heat flux have been determined, it is convenient to define generalized transport coefficients. The relevant elements of the pressure tensor are characterized by an effective (reduced) shear viscosity $\eta^*$ and two independent (reduced) normal stresses $\theta_x$ and $\theta_y$ defined as
\beq
P_{xy}=-\eta^*\frac{p}{\nuhs}\frac{\partial u_x}{\partial y},
\label{Z12}
\eeq
\beq
\frac{P_{xx}}{p}=\theta_x,\quad \frac{P_{yy}}{p}=\theta_y.
\label{Z13}
\eeq
If $d\geq 3$, the other $d-2$ normal stress coefficients $\theta_{z}=\cdots=\theta_{d}$ are related to $\theta_x$ and $\theta_y$ by $\theta_z=(d-\theta_x-\theta_y)/(d-2)$.
The non-zero components of the heat flux $q_y$ and $q_x$ define an effective (reduced) thermal conductivity $\lambda^*$ and a (reduced) cross coefficient $\phi^*$, respectively, by
\beq
q_y=-\lambda^* \frac{d(d+2)}{2(d-1)}\frac{p}{m\nuhs}\frac{\partial T}{\partial y},\quad q_x=\phi^* \frac{d(d+2)}{2(d-1)}\frac{p}{m\nuhs}\frac{\partial T}{\partial y}.
\label{Z14}
\eeq

{}From Eqs.\ \eqref{15}--\eqref{18} we easily get
\beqa
a(\al)&\equiv&\frac{1}{\nuhs}\frac{\partial u_x}{\partial_y}=\left(\frac{1+\al}{2}\right)^2\at\nn
&=&\frac{d+1-\al}{2d}\sqrt{\frac{d+2}{2}(1-\al^2)},
\label{Z22}
\eeqa
\beq
\eta^*(\al)=\left(\frac{d}{d+1-\al}\right)^2,
\label{Z15}
\eeq
\beq
 \theta_y(\al)=\frac{d}{2}\frac{1+\al}{d+1-\al},\quad \theta_x(\al)=d-(d-1)\theta_y(\al),
\label{Z16}
\eeq
where use has been made of Eq.\ \eqref{X6a}. In Eq.\ \eqref{Z22}, $a$ is the shear rate scaled with the collision frequency $\nuhs$. The second equality of Eq.\ \eqref{Z16} implies that $\theta_z=\theta_y$ (if $d\geq 3$).
Note that, from Eqs.\ \eqref{Z4}, \eqref{X6a}, \eqref{Z22}, and \eqref{Z15}, we have
\beq
\eta^*(\al)a^2(\al)=\frac{d}{2}\frac{\zeta(\al)}{\nuhs},
\label{consist}
\eeq
in consistency with $\partial_y q_y=0$ in Eq.\ \eqref{Tbal}.

The heat flux coefficients are given by
\beq
\lambda^*(\al)=\frac{16(d-1)}{d(d+2)(1+\al)^2}\frac{-\mathcal{M}_{2|1,0}}{\et},\quad \phi^*(\al)=\frac{16(d-1)}{d(d+2)(1+\al)^2}\frac{-\mathcal{M}_{2|1,1}}{\et}.
\label{Z20}
\eeq
More explicitly, for $d=3$,
\beq
\lambda^*(\al)=\frac{36 }{  (4 - \al)^2 (829 - 162 \al - 91 \al^2)}\frac{A(\al)}{C(\al)},
\label{Z23}
\eeq
\beq
\phi^*(\al)=\frac{72\sqrt{5/2} \sqrt{1-\al^2}}{ 5(1+\al)  (4 - \al)^2 (829 - 162 \al - 91 \al^2)}\frac{B(\al)}{C(\al)},
\label{Z24}
\eeq
where the functions $A(\al)$, $B(\al)$, and $C(\al)$ are given by Eq.\ \eqref{Z19} and Table \ref{tab:1}.

\section{Comparison with computer simulations}
\label{sec5}

The exact solution to the BE
\eqref{2.1} and \eqref{1} obtained by the moment method in Sec.\ \ref{sec3} defines a normal or
hydrodynamic solution where the spatial dependence of all the moments only occurs
through the hydrodynamic fields $n(y)$, $u_x(y)$, and $T(y)$, as shown by Eqs.\ \eqref{6}, \eqref{3.15}, and \eqref{3.16}. In fact, this solution is free from
boundary-layer effects.

\begin{figure}
\resizebox{0.75\columnwidth}{!}{%
  \includegraphics{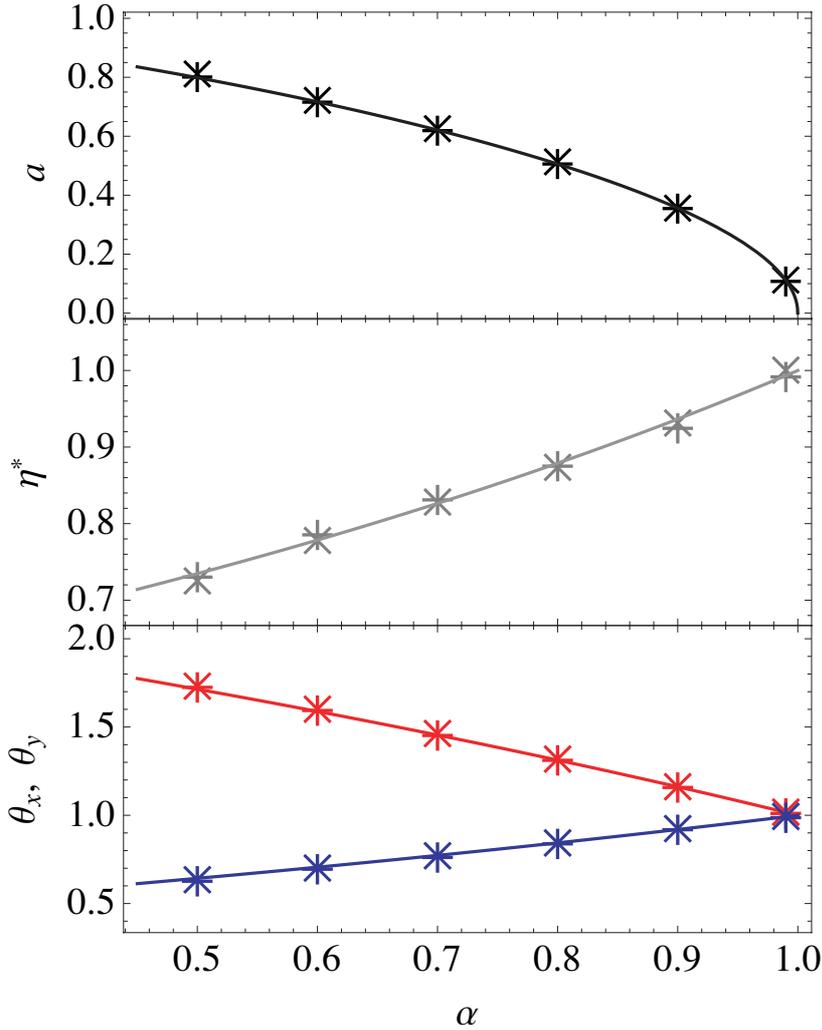} }
\caption{Plot of the reduced shear rate $a$ (top panel), the reduced shear viscosity $\eta^*$ (middle panel), and the reduced normal stresses $\theta_x$ (bottom panel, upper curve) and $\theta_y$
(bottom panel, lower curve) for a three-dimensional system. The lines are the theoretical results \protect\eqref{Z22}--\protect\eqref{Z16}, while the symbols correspond to DSMC results using two different applied thermal gradients: $\Delta T/T_{wd}=2$ ($\times$) and $\Delta T/T_{wd}=5$ ($+$).}
\label{fig1}
\end{figure}

\begin{figure}
\resizebox{0.75\columnwidth}{!}{%
  \includegraphics{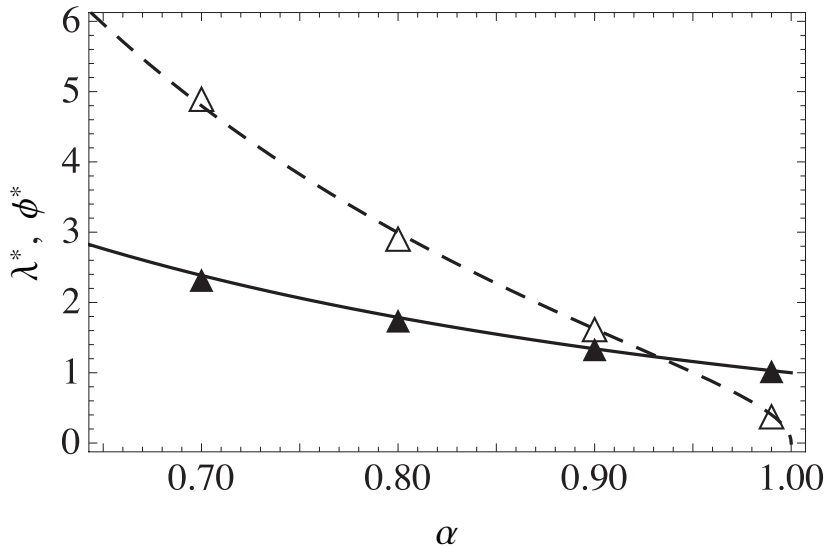} }
\caption{Plot of the reduced heat flux coefficients $\lambda^*$ (solid line and filled triangles) and $\phi^*$ (dashed line and open triangles) for a three-dimensional system. The lines are the theoretical results \protect\eqref{Z23} and \protect\eqref{Z24}, while the symbols correspond to DSMC results using two different applied thermal gradients
($\Delta T/T_{wd} =2$ and $5$).}
\label{fig2}
\end{figure}

The important point is whether or not the exact
solution derived here actually describes the steady state reached by the system,
in the bulk domain, when subject to realistic boundary conditions
and for arbitrary initial conditions. To confirm this expectation,
we have obtained the numerical solution of Eqs.\  \eqref{2.1} and \eqref{1} by means of the DSMC method \cite{B94} for a three-dimensional ($d=3$) system.

The DSMC algorithm consists of two basic steps, both iterated each time interval $\delta t$:  (a) a free streaming step, in which all particles are drifted by $\mathbf{v}\delta t$  ($\mathbf{v}$ being the particle velocity), and (b) a collision step, in which $\nu\delta t$ collision pairs are selected among neighbor particles, where $\nu$ is the characteristic collision frequency appearing in Eq.\ \eqref{1} and given by Eq.\ \eqref{4b} to mimic hard spheres. Additionally, for a bounded system like ours (see Fig.\ \ref{setup}), the particle velocities are updated during step (a) if they eventually touch the boundaries, which in our case are thermal walls moving with constant and opposite velocities (see more details on the description of these boundary conditions elsewhere \cite{VGS08}). The difference between DSMC for the Maxwell model and DSMC for hard spheres lies in the collision step. Since now the collision probability is independent of the relative velocities of the colliding particles [see Eq.\ \eqref{1}],  the collision pairs are randomly selected with equiprobability. Once the particles are selected for collision their velocities are  updated following the same collision rule as for hard spheres [see Eq.\ \eqref{3b}]. The number of particles $N$ and the time step $\delta t$ need to be sufficiently large and small, respectively \cite{B94}. We use here $N=2\times 10^5$  and $\delta t=3\times 10^{-3} \bar{\nu}^{-1}$, where $\bar{\nu}$ is given by Eq.\ \eqref{4b} with $n\to\bar{n}$ and $T\to T_{wd}$, $\bar{n}$ and $T_{wd}$ being the average density and the cold wall temperature, respectively. In addition, the system needs to be split into small enough cells \cite{B94} and we chose layers of width  $\delta y=2\times 10^{-2} \lambda_\text{mfp}$, $\lambda_\text{mfp}=(\pi\sqrt{2}\bar{n}\sigma^2)^{-1}$ being the characteristic mean free path.

In order to find the  states with uniform  heat flux in the bulk, we run series of simulations with varying wall relative speeds $U_{wu}-U_{wd}$ for fixed wall temperature difference
$\Delta T=T_{wu}-T_{wd}$ and distance $h$.
This results in series of varying shear
rate \cite{VU09,VSG09}. Since the theoretical values of the reduced shear rate $a(\al)$ for IMM are very close to those for IHS, we have used values of $h$ and $\Delta T$ in the same ranges as those used in Ref.\ \cite{VSG09}: $h/\lambda_\text{mfp}\sim 10$--$20$, $\Delta T/T_{wd}=1$--$20$; the appropriate range of $U_{wu}-U_{wd}$ depends on the value of $\alpha$ \cite{VSG09}. In order to obtain steady states we need to wait for a sufficiently long computing time. In the case of Maxwell particles this time turns out to be much longer than for hard spheres since the third-order moments are coupled to the fourth-order moments of the USF, which have very long relaxation times \cite{SG07}. Also, for better averaging the steady values of the hydrodynamic profiles and transport coefficients, we average twice: in time, by means of repeated measurements  in the steady state at different uncorrelated instants; and in space, by averaging the small simulation cells to larger hydrodynamic cells, analogously to how we proceeded in a former work \cite{VGS08}. The rest of technical details are the same as in \cite{VGS08}.

Figure \ref{fig1} shows the reduced shear rate $a$, the reduced shear viscosity $\eta^*$, and the reduced normal stress coefficients $\theta_x$ and $\theta_y$ as functions of the coefficient of restitution $\alpha$ for a three-dimensional system. The DSMC results obtained by using two different applied thermal gradients are also plotted. Computer simulations show that, for each value of $\alpha$, the values of  $a(\alpha)$, $\eta^*(\alpha)$ and $\theta_i(\alpha)$ are insensitive to the choice of the thermal gradient. This confirms the independence of the rheological properties and the shear rate on the thermal gradient, including the case of no thermal gradient (USF). We also observe that the agreement between theory and simulation is excellent, even for quite strong values of dissipation. Regarding the influence of dissipation on transport properties, we see that the shear viscosity is significantly affected by dissipation since the deviation of $\eta^*$ from its elastic value ($\eta^*=1$) is quite relevant even for moderate dissipation (say $\alpha \approx 0.9$). It is also important to notice that such a deviation from the value in the elastic limit has a  sign opposite to that of the NS  shear viscosity \cite{S03}. With respect to the normal stress coefficients, Fig.\ \ref{fig1} shows that the anisotropy induced by the shear flow is more important in the direction parallel to the flow velocity ($x$ direction) than in the direction parallel to the thermal gradient ($y$ direction).

The $\alpha$-dependence of the (reduced) heat flux coefficients $\lambda^*$ and $\phi^*$ is plotted in Fig.\ \ref{fig2}. As said before, given that the transient regime needed to reach steady-state values of these coefficients is longer than that of the rheological properties (especially for strong dissipation), the simulation data reported here ($\alpha \geq 0.7$)  cover a range of values of $\alpha$ smaller than the one plotted in Fig.\ \ref{fig1}. However,  $\alpha \geq 0.7$ is still of course a relevant range of values of dissipation since most of the experiments carried out for granular fluids are included in this region. Figure \ref{fig2} clearly shows again that the heat transport is clearly affected by collisional dissipation, especially in the case of the cross-coefficient $\phi^*$. As happens for the rheological transport coefficients, the theoretical results compare very well with simulations in all the range of values of $\alpha$ studied. Interestingly, it can be be observed that the streamwise component $q_x$ becomes larger in magnitude than the crosswise component $q_y$ for $\alpha \lesssim 0.9$, what represents a strong far-from-equilibrium effect.

\section{Summary and Discussion}
\label{sec6}

Exact solutions in kinetic theory for far-from-equilibrium states are exceedingly rare. The difficulties increase in the case of a granular gas of smooth IHS since a new parameter (the coefficient of normal restitution $\alpha$) is introduced to account for the dissipative character of collisions. Usually, in order to get explicit results from the BE one considers approximate methods such as the leading order in a Sonine polynomial expansion of the velocity distribution function or the well-known Grad's 13-moment method. Although in many of the cases the results obtained from the above analytical methods compare reasonably well with numerical solutions of the BE, the lack of exact solutions motivates the search for alternative routes. One possibility consists of replacing the detailed Boltzmann collision operator by a simpler collision term (e.g., the Bhatnagar--Gross--Krook model kinetic equation), that otherwise retains the most relevant features of the true collision operator. Another different approach consists of retaining the mathematical structure of the Boltzmann operator, but considering a simplified collision model, such as the IMM. As in the case of elastic collisions \cite{GS03,S09}, the collision rate of this model is assumed to be independent of the relative velocity of the two particles that are about to collide.

The simplicity of IMM allows us, for instance, to get exactly the collisional moments of the BE without the explicit knowledge of the velocity distribution function \cite{GS07}. Their knowledge has opened up the possibility of exactly determining the nonlinear transport coefficients in some far-from-equilibrium situations, such as the USF problem  \cite{G03,SG07}. In this paper, an exact solution for the class of steady Couette flows defined by a uniform heat flux has been found. This solution is characterized by a constant pressure $p=nT$ and by linear profiles of
the $x$-component of the flow velocity $u_x$ and the temperature $T$ with respect to a conveniently scaled space variable. As a consequence, a linear relationship between $T$ and $u_x$ holds, so that we refer to this class of steady Couette flows as the LTu class. Moreover, as in the USF state, the condition of stationarity imposes a relationship between the shear rate and the coefficient of restitution, so that the (reduced) thermal gradient $\et$ is the only free nonequilibrium parameter of the problem. In particular, this steady Couette flow reduces to the USF when $\et =0$ (with $\alpha \neq 1$), while it reduces to the conventional Fourier flow problem when $\alpha=1$ (with $\et \neq 0$).
A previous study  of the LTu flow problem  has been recently carried out by the present authors  for IHS \cite{VSG09} by considering three complementary routes: an approximate solution from Grad's 13-moment method, DSMC simulations, and molecular dynamics simulations.

A careful inspection of the moment hierarchy \eqref{13} (seen as an infinite set of first-order differential equations with respect to $\et$) shows that it can be exactly solved via a recursive scheme. The solution is characterized by reduced moments of the form (\ref{14}), i.e.,   moments of order $k$ are polynomials of degree $k-2$ in the (reduced) thermal gradient $\et$. The coefficients of these polynomials are nonlinear functions of the coefficient of restitution $\alpha$ that can be recursively solved starting from the USF values (where the hierarchical character of the moment equations is broken down) and following the scheme sketched in Fig.\ \ref{arrows}. In particular, the elements of the pressure tensor $P_{ij}$ are independent of $\et$ and their expressions coincide with those of the USF \cite{SG07}. The above elements define an effective (reduced) shear viscosity $\eta^*$ and two independent (reduced) normal stresses $\theta_x$ and $\theta_y$.  The heat flux ${\bf q}$ is linear in $\et$ and has two nonzero components: one component $q_y$ parallel to the thermal gradient and a component $q_x$ normal to the thermal gradient. In reduced units, the first component defines an effective thermal conductivity $\lambda^*$ while the latter component defines a cross-coefficient $\phi^*$ that is absent in the NS description. The evaluation of the explicit dependence of $\lambda^*$ and $\phi^*$ on $\alpha$ has been one of the main goals of this paper. This dependence has been obtained thanks to the knowledge of the fourth-degree velocity moments of IMM in the USF \cite{SG07}. As expected, the results show that the influence of collisional dissipation on the transport coefficients $\eta^*$, $\theta_i$, $\lambda^*$, and $\phi^*$ is quite significant (see Figs.\ \ref{fig1} and \ref{fig2}).
This is especially apparent in the case of $\phi^*$ (which is zero in the elastic limit) since its magnitude becomes even larger than that of the effective thermal conductivity $\lambda^*$
for moderate values of dissipation. This is a signal of the strong non-Newtonian behavior of the LTu flows for the granular gas.

In order to validate the theoretical results derived from the moment method, computer simulations based on the DSMC method have been carried out for the same interaction model in the three-dimensional case. Given that the exact solution derived in Section \ref{sec4} is free from boundary-layer effects, the important point to elucidate via  comparison with computer simulations is whether or not this analytical solution actually describes the steady state reached by the gas in the bulk domain, when the gas is subject to realistic boundary conditions and for arbitrary initial conditions. The good agreement between theory and simulation found here confirms the hydrodynamic profiles and the nonlinear transport properties predicted by the analytical solution, even for quite strong values of dissipation.

\begin{acknowledgement}
This work has been supported by the Ministerio de Educaci\'on y Ciencia (Spain) through grant No.\ FIS2007-60977, partially financed by FEDER funds and by the Junta de Extremadura (Spain) through Grant No.\ GRU09038.
\end{acknowledgement}


\begin{thebibliography}{99}
\bibitem{note}
An independent calculation of the fourth-order moment $\int\dd\mathbf{V}\, V^4 J[\mathbf{V}|f,f]$ was carried out in the Appendix of Ref.\ \protect\cite{BC07}.
\end{thebibliography}
\end{document}